\documentclass[conference]{IEEEtran}
\ifCLASSINFOpdf
\else
\fi
\usepackage{todonotes}
\usepackage{booktabs}
\usepackage{listings}
\usepackage{url}
\usepackage{framed}
\usepackage{multirow}
\usepackage{paralist}
\usepackage{color,soul}
\usepackage{balance}

\newcommand{\highlightt}[2]{\begin{leftbar}\noindent \textsf{\textbf{#1}} \emph{#2}\end{leftbar}}

\begin{document}
\title{On the Relationship of Inconsistent Software Clones and Faults: An Empirical Study}

\author{\IEEEauthorblockN{Stefan Wagner, Asim Abdulkhaleq and Kamer Kaya}
\IEEEauthorblockA{Institute of Software Technology, \\ University of Stuttgart, Germany\\
  \\
 }
\and
\IEEEauthorblockN{Alexander Paar}
\IEEEauthorblockA{TWT GmbH Science \& Innovation\\
Stuttgart, Germany 
 }}


\maketitle

\begin{abstract}
\emph{Background:}
Code cloning -- copying and reusing pieces of source code -- is a common phenomenon in
software development in practice. There have been several empirical studies on the effects
of cloning, but there are contradictory results regarding the connection of cloning and faults.

\emph{Objective:}
Our aim is to clarify the relationship between code clones and faults. In particular, we 
focus on inconsistent (or type-3) clones in this work.

\emph{Method:}
We conducted a case study with TWT GmbH where we detected the code clones in three Java systems,
set them into relation to information from issue tracking and version control and interviewed three key developers. 

\emph{Results:}
Of the type-3 clones, 17~\% contain faults. Developers modified most of the type-3 clones simultaneously
and thereby fixed half of the faults in type-3 clones consistently. Type-2 clones with faults all
evolved to fixed type-3 clones. Clone length is only weakly correlated with faultiness.
 
\emph{Conclusion:} 
There are indications that the developers in two cases have been aware of clones. It might be a 
reason for the weak relationship between type-3 clones and faults.
Hence, it seems important to keep developers aware of clones, potentially with new tool support.
Future studies need to investigate
if the rate of faults in type-3 clones justifies using them as cues in defect detection.
 \end{abstract}


% \fi
\IEEEpeerreviewmaketitle

 \section{Introduction} 

Hunt and Thomas~\cite{hunt99} introduced the DRY principle in software engineering: Don't 
repeat yourself. They wanted to reduce repetition in software systems in all areas, especially source code.
The reason was that a modification of one source code element should not cause further changes in other
elements. We commonly call the results of violating the DRY principle \emph{clones} and the actions
leading to it \emph{cloning}.

The research community has, over time, come up with a plethora of different techniques and
tools to detect clones in source code~\cite{Krinke:2001ev,Kamiya2002,Deckard,Juergens:2009hy,Komondoor:2001it,Kim:2011jh} 
but also other artefacts~\cite{pham2009complete,Deissenboeck:2008hm,Juergens:2010:CDS:1810295.1810308}.
It is now possible to reliably and precisely detect most types of code clones automatically. We
often see rates of 20~\% to 30~\% of a system to be clones in practice~\cite{wagner2013software}.

One specific area of interest in research have been clones with inconsistencies between them. In particular,
these clones underwent changes that go beyond renaming and layout modifications: Statements have been changed,
added or deleted. We call such clones \emph{inconsistent}, \emph{gapped} or \emph{type-3} clones. These
clones are especially interesting since their inconsistencies could be the result of incomplete fixes, i.e.\ a bug
report could have led to a fix in one clone instance but not in the others.

\subsection{Problem Statement}
Yet, it is still not clear what effects (negative and positive) cloning really has. There is an ongoing debate
about the harmfulness of cloning especially in source code. On the one side, Kapser and Godfrey~\cite{Kapser:2008en} claim that 
there are many useful patterns of cloning in software systems. Rahman, Bird and Devanbu~\cite{rahman10} found
that ``the great majority of bugs are not significantly associated with clones.'' On the other side, for example, Juergens et al.~\cite{Juergens:2009hy}
found 107 faults associated with the inconsistencies between clones in five professional software systems. So what
are the impacts of cloning?

\subsection{Research Objectives}
Our overall goal is to understand the effects and implications of cloning in software artefacts. In this paper,
we concentrate on the effects of type-3/inconsistent clones on faults in software and the context in which
these effects occur.

\subsection{Contribution}
We contribute an investigation of the relationship between type-3 clones and faults in an industrial case
study. We analyse three closed-source software systems written in Java over their whole life time. We make use
of the history in the version control and issue tracking systems to
\begin{itemize}
\item quantify the share of type-3 clones overall,
\item quantify the share of faulty type-3 clones,
\item analyse the awareness of developers of clones and
\item quantify the correlation between clone length and faults.
\end{itemize}
Furthermore, we investigate the project lead of each investigated system to triangulate the results
on awareness with the view of the developers in these projects.

\subsection{Context}
We performed the case study at the German company TWT GmbH which offers engineering and IT services in the sectors
automotive, aerospace, healthcare and energy. TWT has around 300 employees in six local offices. In this study, we investigated
projects and systems in the automotive domain with an age of 4--5 years written in Java.

The reporting of this study follows largely Runeson and H\"ost's \cite{runeson2009guidelines} guidelines.

\section{Terminology}
The clone detection community has established a common terminology to talk about clone-related concepts.
We follow this terminology based on existing surveys and reviews~ \cite{koschke-2007_KoschkeR-clone_survey,Rattan:2013hm}
in our paper. We summarise the most relevant terms in the following:

\textbf{Code fragment:} A \emph{code fragment} is a sequence of statements including control statements such as loops. A code
fragment can be any contiguous sequence of lines in a file.

\textbf{Code clone:} One code fragment is a \emph{clone} of another code fragment if they are similar according to a given
definition of similarity. There are different such definitions forming different \emph{clone types}.

\textbf{Clone group:} A set of clones that are all similar to each other is called a \emph{clone group} (also called \emph{clone class}).

\textbf{Clone instance:} Sometimes it is useful to be able to talk about an individual member of a clone group. We call this
member a \emph{clone instance}.

\textbf{Type-1 clone:} A type-1 clone is a fully identical code fragment ``without modification (except for white space and comments).''~\cite{4288192}

\textbf{Type-2 clone:} For type-2 clones, we allow more variations. A type-2 clone is ``a syntactically identical copy; only variable, type, or function
identifiers were changed.''~\cite{4288192}

\textbf{Type-3 clone:} A type-3 clone is a type-2 clone ``with further modifications; statements were changed, added, or removed.''~\cite{4288192}
Type-3 clones are also called \emph{inconsistent clones}, \emph{near-miss clones} or \emph{gapped clones}.

Furthermore, as we investigate the relation of type-3 clones to faults, we need to define what a fault is exactly.
We first define \emph{failure} and use that to define \emph{fault}.

\textbf{Failure:} ``A failure is an incorrect output of a software visible to the user.''~\cite{wagner2013software}

\textbf{Fault:} ``A fault is the cause of a potential failure inside the code or other artefacts.''~\cite{wagner2013software}

\section{Related Work}

We structure the related work into clone-based fault detection techniques, empirical studies
of clone genealogies involving inconsistencies between clones and manual inspections of
type-3 clones.

\subsection{Clone-Based Fault Detection}
As cloning was considered harmful including a higher potential for faults, in the mid-2000s, approaches
to detect faults based on clone analysis were proposed. Most notably, there were two proposals of such
approaches that the researchers validated in empirical studies:

Li et al.~\cite{Li:2006ip} proposed to detect faults based on inconsistent renaming in code clones. They
tested their approach on Linux and FreeBSD where they were able to find 49 and 31 faults, respectively.

Jiang, Su and Chiu~\cite{jiang2007context} developed an approach for fault discovery using context-based
inconsistencies between clones. They validated their approach on the Linux kernel and Eclipse. They also
were able to detect a number of faults: 57 in the Linux kernel and 38 in Eclipse.

Hence, there seem to be faults in real software systems associated with inconsistencies between clones. Our
approach, however, focuses on documented faults instead of detecting new faults.

\subsection{Clone Genealogies}

Kim et al.~\cite{Kim:2005:ESC:1095430.1081737} introduced the notion of \emph{clone genealogies}: ``The
genealogy of code clones describes how groups of code clones change over multiple versions of a program.''
This kind of approach was used in several studies to investigate if type-3 clones are associated to faults.

Krinke~\cite{Krinke:2007gv} analysed five open-source systems and their clone genealogies for inconsistencies
between clones. He found that half of the changes in the systems' history were inconsistent. Yet, inconsistent
clones did evolve mostly independently which is an indication that they did not represent faults.

Bakota, Ferenc and Gyimothy~\cite{Bakota:2007bj} also followed the evolution of clones of the history
of an open-source system: Firefox. They found two faults that had been documented and fixed as well as
five further potential faults.

Thummalapenta et al.~\cite{Thummalapenta:2010bv} investigated clone genealogies in four open-source
systems. They found that below 16~\% of clone groups undergo late propagation of changes. They interpret
their results such that these late propagation are likely to be caused by bug fixes. 

Rahman, Bird and Devanbu~\cite{rahman10} analysed clone genealogies in four open-source projects. They
found that the majority of reported defects in the projects were not associated with clones. They also found that
clones are less defect-prone and faults in clones need a similar fix effort as in other code.

Barbour, Khomh and Zou~\cite{barbour13} especially investigated faulty clone genealogies and different patterns of
inconsistent clones. They found that certain types of late propagations between clones are most risky
to be faulty.

Mondal, Roy and Schneider~\cite{mondal15} compared the fault proneness of different types of clones in a set of open-source
projects. They found that type-3 clones are the most fault-prone.

Type-3 clones seem to be a significant part of all clones, and there are faults associated with these clones.
They do not seem to be a major source of faults overall, however.

\subsection{Manual Inspection}

Juergens et al.~\cite{Juergens:2009hy} is the only study known to us that analysed closed-source systems so far.
They investigated four industrial systems from two companies and an academic open-source system for type-3 clones.
The researchers showed all type-3 clones, which they considered to be true positives, to the developers of the
systems and let them annotate them whether the inconsistency was intentional and it constitutes a fault. They found
that 28~\% of type-3 clone groups had unintentional inconsistencies and of these every second was a fault.

G\"ode and Koschke~\cite{Gode:2011ct} investigated the clone genealogies in three open-source systems for
changes to clones and unintentional inconsistent changes. They found that 57~\% of changes were consistent.
They inspected the remaining changes manually. They found that only 15~\% of changes seemed to be unintentionally
inconsistent.

Bettenberg et al.~\cite{Bettenburg:2012fs} investigated clone evolution in three open-source systems. They
manually inspected all inconsistent changes themselves and judged whether the change should have been
applied to all clone instances, i.e.\ introducing a fault. They found that of these inconsistent changes 1~\%--4~\%
introduced faults.

\subsection{Summary}
Type-3 clones seem to be a significant phenomenon in real software systems. Various studies found faults
by investigating these type-3 clones. There have also been indications that type-3 clones often are not associated
to faults. Overall, type-3 clones do not seem to be more fault-prone than other code. Several of the studies rely on
manual inspections of the developers or even the researchers. This is a validity risk. Even in the case of developer
inspection, it raises the question whether these faults are interesting as they have not caused issue reports so far.
Hence, we concentrate on the relationship of type-3 clones with documented faults.

Furthermore, an overall weakness of the empirical body of knowledge on type-3 clones and faults is that almost 
all studies done so far concentrate on open-source software. We investigate the repositories of three industrial 
closed-source software systems and have access to key developers of the systems.

 \section{Case Study Design}
 Our case study is descriptive as it describes the situation of type-3 clones and faults in industrial closed-source software 
 as well as explanatory as we
 investigate factors influencing the relationship between type-3 clones and faults. But as we have a
 post-ex-facto design, our results are mostly correlational. 

\subsection{Research Questions}
The goal of this research is to investigate the relationship between type-3 clones and software faults. We formulate three
main research questions to structure our analysis for this study goal. 

\noindent\textbf{RQ~1: Do software systems contain type-3 clones?}
 
Earlier studies (e.g.\ \cite{Juergens:2009hy}) have shown that there is a considerable share of type-3 clones in
relation to all clones. If type-3 clones were rare, it would not be interesting to further investigate their relationship to faults. 
With this question, we want to validate the earlier findings.

\noindent\textbf{RQ~2: Do type-3 clones contain documented faults?}

If we find a considerable number of type-3 clones in the cases, we can answer a main aspect of our research goal
by checking the correspondence between documented faults in the analysed systems and their occurrence in type-3
clones. We are interested in whether there are documented faults at all and, if so, the share of type-3 clones with such faults.

\noindent\textbf{RQ~3: Are developers aware of type-3 clones?}

From previous studies (e.g.\ \cite{Juergens:2009hy,Thummalapenta:2010bv}), we expect that the faultiness of
type-3 clones is influenced by whether developers introduce inconsistencies intentionally or unintentionally.
In other words: Are the developers aware that there are clones and that they introduce an inconsistency to them?
As we could not observe the developers during the changes and had access to only a subset of them directly, we 
use four sub-research questions to investigate direct and indirect indications of awareness.

\noindent\textbf{RQ~3.1: Do developers maintain type-3 clones simultaneously?}

Our first indication of the awareness of developers of clones and especially type-3 clones is whether changes to
these clones are done simultaneously. If during a change involving one instance of a type-3 clone group also
the other instances are changed, it is likely that the developers are aware of the clone group.

\noindent\textbf{RQ~3.2: Are faults in type-3 clones fixed consistently in all instances of a clone group?}

The second indication of awareness are consistent fixes of faults in type-3 clones at all clone instances. The
developers then had to be aware of the clones and the need to fix the fault in all of the instances.

\noindent\textbf{RQ~3.3: To what degree do type-2 clones with a documented fault become type-3 clones without a documented fault?}

The third indication of developer awareness of clones is whether type-3 clones could simply be the result of
successful fault fixes in type-1 or type-2 clones. A fault fix in a clone instance may create a type-3 clone. Nevertheless,
it is possible that the fix is only necessary in one instance. Hence, we want to understand whether this scenario
happens in real systems.

\noindent\textbf{RQ~3.4: How do the developers deal with clones?}

The last part of the awareness question are statements from key developers of the investigated projects. We
want to understand how the developers perceived how they dealt with clones overall and during bug fixes and
other changes. This completes our overall view on the awareness of clones.

\noindent\textbf{RQ~4: Are longer type-3 clones more likely to contain faults?}

Finally, we want to investigate another context factor: the length of clones. Intuitively, longer clones should 
contain more faults just because they also contain more statements that can be faulty. We are interested
in whether the clone length has an influence on the faultiness of clones.

\subsection{Case Selection}
The main consideration for selecting cases in this case study was availability: We selected cases to which we had access via the 
collaboration with an industry partner to collect the necessary data. Furthermore, we selected cases with a size of at least
200,000 LOC and an age of at least four years. We considered that to be necessary to have enough clone instances for a reasonable
analysis.

\subsection{Data Collection Procedure}
For the analysis, we needed data from version control, clone detection, issue management tools and developers from
the three projects. We performed the collection of data in six steps:

\subsubsection{Extraction of Latest Source Code Version}
As there is considerable manual work in our further analysis procedures, we refrained from running a full genealogy analysis.
Instead, we chose to detect clones in the latest versions of the source code for each system. From the detected clones in these
versions, we will work backwards and analyse the history of the files with clones.

The cases we investigated used the version control system \textbf{Mercurial}~\cite{o2009mercurial}. It is a platform-independent, 
distributed version control system. It handles small and large projects and is widely used in practice.

In the cases, the developers used in addition the Web-based system \textbf{Kiln}.\footnote{\url{http://www.fogcreek.com/kiln/features/team-up/}} 
It hosts source code with the version control systems Git and Mercurial. When using Mercurial or Git, the Kiln server is the central point of the 
version control system. The data are stored centrally. In other words, Kiln has a dual function. It serves as data storage and as a distribution 
node for the source code as well as other project-related files. This supports distributed software development teams.

From Kiln and Mercurial, we extracted the latest versions of all the source files for each system.

\subsubsection{Clone Detection} 
Next, we need the clones in these latest source code versions. There are several tools available to perform clone detection. We
chose the tool \emph{ConQAT} because it allows us a type-3 clone detection, it is recommended in \cite{ICSME-2014-SvajlenkoR}
and we had previous experience with it. 
The Continuous Quality Assessment Toolkit (\emph{ConQAT}) \cite{4602675} is a general toolkit for continuous software quality control and
analysis. It supports rapid development and execution of software quality analyses, for example architecture conformance analysis. 
ConQAT has been developed since 2007 at the Technische Universit\"at M\"unchen and has now commercial support by CQSE GmbH.
It is still available as open-source software. ConQAT itself is based on a pipes-and-filters architecture and offers a data-flow language
for the specification and parameterisation of the clone detection pipeline. ConQAT has been used for clone assessments 
in several studies \cite{Juergens:2010:CDS:1810295.1810308, Juergens:2009hy} and is also applied in industry.

In particular, we used the ConQAT block \emph{JavaGappedCloneAnalysis.cqr} to detect type-3 clones. The algorithm for detecting type-3 was 
developed by Juergens et. al.~\cite{Juergens:2009hy}. As the parameter settings for the clone detection can have an effect on the outcome,
we ran the analysis with two different settings. We performed the detection once with conservative and once with liberal clone detection parameters. The liberal parameter setting was a min-length of 10, max.\ of errors of 10 and ratio of gap of 0.25. For the conservative setting, we used a min-length of 20, max.\ of errors of 10 and gap ratio of 0.25. 
Finally, we used the conservative approach for the further analysis. This gives us the lists of clone groups, clone instances and files with clones.
Wang et al.~\cite{conf/sigsoft/WangHJK13} proposed an approach to systematically search the configuration space of
clone detection tools. We consider this extensive approach not necessary in a case study, however.

\subsubsection{Extraction of all Revisions of Files with Clones} 
As we aimed to investigate the entire revision history of the investigated systems, we needed to extract for all files with clones the corresponding
earlier versions. We could again use Mercurial for this job. We extracted the paths of files containing clones from the ConQAT clone detection results 
and stored them in a separate text file. We wrote a Python script that created another text file for each entry of the revision history from Mercurial for
each file with clones. The version history consists of change sets which are caused by commits. For each change set, Mercurial creates a local and 
a unique \emph{ChangesetID}. In addition, it contains the user who has triggered the commit and thereby created the change set, the date of the 
commit, its branch, its parent commit and a description of the commit. For the further analysis, we particularly needed the \emph{ChangesetID}, the
\emph{user}, the \emph{date} and the \emph{description of change sets}. We executed the script on the command line for each system.

The output of the script was exported into a Microsoft Excel file and rewritten with Excel references and functions. The output also lists each file that was recorded in 
the previous text file and the change set with the above information. With this information, we can check out any relevant version of the files with clones.

\subsubsection{Extraction of Issues for Files with Clones} 
We now have all versions of files that contained clones in the latest version. Next, we need to collect data to establish the connection to faults. For that, we
extract data from issue tracking. In our cases, the developers used \emph{FogBugz} for project management and issue tracking. 

\textbf{FogBugz} is a project-management and issue-tracking system which offers broad functionality for development teams. The issue tracking allows users 
to manage, filter, sort and navigate a tree-structure of tasks that contain information, tags and attached files related to a particular issue. FogBugz tracks all 
events and tickets in one central location~\cite{FogBugz}. In particular, these tickets can describe faults.

The history of the files with clones consists of \emph{ChangesetIDs} with additional information as explained above. To match the faults, we first determined 
the type of clone, i.e.\ whether it is an inconsistent or a consistent clone. Faults that have occurred at some point in the development or usage of the system
are documented as issues in FogBugz and categorised as faults. Each issue has a unique issue number. If a source code modification resulted from an issue,
the developers include the issue number in the commit message as a reference. In summary, this means we need to check \emph{ChangesetIDs} which
resulted from commits which reference faults. In the end, the list of revision histories was extended with the referenced issue numbers. Through this, all files 
with clones that have been modified to fix a fault were identified. 

\subsubsection{Collection of Meta-Data in a Database}
After performing the clone detection with ConQAT on the systems under analysis and extracting the data from the systems' repositories, we established the structure 
of a Microsoft Access database to store all data. This simplified the analysis procedure. The structure of the database contains the following items: 

\begin{itemize}
\item A data table (``clonegroup'') that includes the clone groups which are derived from the results of the inconsistent clone detection with ConQAT.
\item A data table (``clonefile'') that stores the clone file names of clone groups which are derived from the ConQAT results.
\item A data table (``history'') that contains the extracted data from FogBugz and Kiln which includes all clone files with their entire version history. The version history 
consists of change sets with information such as user, date, summary and clone file identification and the issues which describe the faults.
\item A relation data table (``relation'') that contains the relationships between the three other tables.  
 \end{itemize}

\subsubsection{Interviews with Developers}
The last step in the data collection was to conduct interviews with developers of all investigated
systems. We needed at least one developer from each case who has at least several months
of experience in developing the system. Ideally, they should be in a key role, such as the project
lead, who has a good overview of how the projects are conducted.

We conducted semi-structured interviews over telephone by one of the researchers. Each interview
was digitally recorded and then transcribed (directly in the tool \emph{MaxQDA}) by the same researcher. We
took care to emphasise that we are mostly interested in the experiences of the developers, not their
opinion. Only in the question of a potential IDE plugin showing clones while developing, we asked
for opinions in case they do not already use one. The interviewer used the following guiding questions:

\begin{compactenum}
\item For how long have you worked in software development? With what programming languages and technologies?
\item How are you involved in which project?
\item Have you been aware of the concept of code clones in general?
\item If you fixed a fault in the system, did you explicitly look for clones?
\item How would you describe your (and your colleagues') awareness of clones in your project?
\item Have some sort of clone detection tools been used in the project?
\item How useful is/would be an IDE plugin showing clones?
\end{compactenum}

\subsection{Analysis Procedure}
To analyse the data which we stored in the database, we created SQL queries for answering the research questions. We performed the 
following steps to get the necessary data for each research question.

\subsubsection{Type-3 Clones (RQ~1)}
We first need to extract the numbers of all clone groups and the numbers of type-3 clone groups to set them into relation.

We run a query on our database which returns all type-3 clone groups. We determine the number of type-3 clone groups $C^{T3}$
by counting the first occurrence of each type-3 clone group in the analysed versions from the repository.
Setting $C^{T3}$ into relation with the number of all clone groups $C$ will answer RQ~1.

\subsubsection{Faulty Type-3 Clones (RQ~2)}
Second, we retrieve the entire revision history of the file list from the database. Then, we extract the clone groups which have a fault by using a second SQL query.
By this query, all clone groups (consistent and inconsistent) $C_F$ are determined in which a fault has been resolved. We will use all faulty
clone groups in a later analysis. With a similar query, we get only the faulty type-3 clone groups $C_F^{T3}$. By setting the number
of the faulty clone groups into relation to all clone groups we can answer RQ~2.

\subsubsection{Simultaneous Maintenance (RQ~3.1)}
For this analysis, we extracted all instances of type-3 clones $I$. We inspected them and extracted the clone instances that
have been modified over the revision history. This gives us $I_M$.

For all those modified instances, we manually checked if they were simultaneously modified (in the same commit) with the other clones in the same
clone group. We call these $I_{MS}$. When we set them into relation to $I_M$, we can answer RQ~3.1.

\subsubsection{Consistent Fixes (RQ~3.2)}
Furthermore, we inspected each change to a type-3 clone related to a fault and checked whether the same fix was
applied consistently to all clone instances. If that was the case, we count them under $C_X$. Setting $C_X$ into relation
to $C_F^{T3}$ answers RQ~3.2.

\subsubsection{Faulty Type-2 to Non-Faulty Type-3 (RQ~3.3)}
Moreover, we extracted from our database also all faulty type-2 clone groups (as described above). These clone groups
are called $C_F^{T2}$. From each of these clone groups, we went forward in the revision history and looked for fixes.
If such a fix occurred, we checked whether this resulted in a type-3 clone group and if there is a further fix of this clone
group in the future. We describe the faulty type-2 clones that evolved into non-faulty type-3 clones with 
$C^\mathit{T2}_F \rightarrow C^\mathit{T3}_\mathit{NF}$. Setting them into relation to all faulty type-2 clones
answers RQ~3.3.

\subsubsection{Developers' View (RQ~3.4)}
We employed a light-weight coding approach for the qualitative analysis of the interviews. As we did not aim at 
establishing a broader, grounded theory of cloning, we concentrated on creating codes (tags) that relate directly to our
research question on the developers' awareness of clones in their software system. Yet, we also added codes
if they seemed interesting to describe the context of the case or the wishes for improvement by the developers.

Using the codes, we analyse the differences between the projects, describe them and set them into relation to
the results of the quantitative analyses we performed for the RQs above. The aim is to check if we can derive
a consistent understanding of how aware the developers were of clones in the development of their systems.

\subsubsection{Clone Length and Faultiness (RQ~4)}
We can extract the length of a clone in units directly from the ConQAT results. We use a correlation analysis using 
Spearman's rho for investigating the relationship between faultiness of clones (0 or 1) and the length in
statements of clone groups. We define the following null hypothesis:

\emph{H0. There is no difference in the length of clones between clone groups with a fault and clone groups without a fault.}

and the corresponding alternative hypothesis:

\emph{H1. There is a difference in the length of clones between clone groups with a fault and clone groups without a fault.}

In case there is a statistical relationship, we investigate this hypotheses with t-test. The results of the statistical tests 
determine RQ~4.

\subsection{Validity Procedure}
 
\subsubsection{Construct validity} 
The development history of the systems was analysed. A problem was that code fragments can be inserted by copying and be modified 
in a single commit. Therefore, we manually inspected the entire revision history of the systems under consideration to check all changes 
of a code fragment. Another threat to construct validity is that the issues referenced in a commit might not fit to the actual change. To counter 
this threat, we examined the code for each issue to see whether the changes corresponded to it.

\subsubsection{Internal validity}
The clone detection process can yield false positives and false negatives. As we used a tool to detect inconsistent clones in the three systems, we manually
inspected all found clones for false positive results.  Another internal threat to validity is the configuration of ConQAT. We carried out the 
clone detection with a liberal and conservative approach in the configurations to check its influence. 

The qualitative analysis by codes is inherently subjective to some degree. We used a review of the codes produced by one
researcher by two of the other authors. Potential misunderstandings can be discussed and resolved this way.

\subsubsection{External Validity} 
A possible external threat to validity is that the systems do not represent software systems and software development
at large. We chose our selection criteria to reduce this threat.

\section{Results}

 \subsection{Case Description}
This study was carried out on systems of the German company \emph{TWT GmbH}. The company offers engineering and IT services in the sectors
automotive, aerospace, healthcare and energy. It has around 300 employees in six local offices. 

According to our case selection criteria, we chose three software systems which are in development and have a reasonably long development history. 
These systems are developed in Java by different teams and offer different functionalities. Moreover, the systems are already in live operation and are continuously 
developed further and adapted. For privacy reasons, the names of the systems are anonymised as A, B and C. As detailed in Tab.~\ref{tab:studyobjects}, the systems have been
in development for 4 to 5 years, have a size between 250 and 450 KLOC and had 5 to 10 developers.

 \begin{table}[htb]
  \caption{Summary of the Cases}
  \centering
 \begin{tabular}{l l l r r r r}
  &  &  & Size &  & Age &   \\
 System & Domain & Lang. & (KLOC) & Revisions & (Years) & Developers  \\
\toprule
A     & Automotive  & Java  & 253   & 2470  & 4     & 10  \\
B     & Automotive  & Java  & 332   & 1622  & 5     & 5  \\
C     & Automotive  & Java  & 454   & 2181  & 4     & 10  \\
\bottomrule
 \end{tabular}%
   \label{tab:studyobjects}
\end{table}

All three software systems are Java rich client applications based on Eclipse RCP and an Oracle relational database backend. All systems have been developed by similar developer teams concerning age, academic education and practical experience. Also, all systems have a similarly sized user base. The users of all three systems work in digital development processes in the automobile industry.

System A facilitates the management of sensors, test campaigns, measured data and simulation results in the area of digital verification. In particular, the software comprises a variety of interfaces for the import/export of data files. Test campaigns to collect measured data are planned to complete and to validate simulation data where necessary. For this purpose, system A contains seven different Eclipse application perspectives with different editors and views each.

Systems B and C are information systems for $\mbox{CO}_2$ emission data and related technological measures. System B provides this functionality for individual car configurations. In system C, type series are first class citizens. Due to their functional similarity, during the initial development phase of system B, a significant portion of its code base was derived from system C. In contrast to system A, systems B and C do only contain one Eclipse application perspective that contains the complete set of editors and views. Moreover, these controls are coupled to facilitate, for example, facetted browsing.

All three systems make use of a TWT proprietary class library (TFC) for model-driven software development. The relational data model is therefore transformed by means of the Hibernate tools into Java classes for domain objects, single- and collection properties. Property classes allow for genericity (e.g.\ one can iterate over all properties of an object without knowing its properties by name). Additional metadata and Java annotations facilitate generic controls such as, for example, edit dialogs for master data objects. In the TFC and all three systems, UI controls are implemented based on SWT and JFace.

For all three systems, the TFC is referenced as a Mercurial sub-repository. Thus, the library is developed further with direct commits from the application specific software development projects (i.e. there is no baseline configuration management for the TFC). Modifications of the TFC are synced up in biweekly developer meetings. The initial development of the TFC started when a predecessor of system A was then ported from a previous code base to Eclipse RCP.

We were able to interview the project leads for all three cases. All three of them have a university degree
in computer science, two of them even a PhD. All three have several years of experience in Java and other
programming languages.

\subsection{Share of Type-3 Clones (RQ~1)} 

Table~\ref{tab:results} contains the quantitative results for all research questions in detail.
We found a mean share of type-3 clones in all clones and all three systems of 52~\%. Yet,
it varied strongly from 23~\% in system B to 79~\% in system C. Nevertheless, in all
three systems, there is a considerable share of type-3 clones and, hence, it is useful to
investigate their relationship with faults.

\begin{table}[htb]
  \caption{Summary of results}
  \centering
  \begin{tabular}{p{4.5cm} r r r r}
    \toprule
    Project & A & B & C & Total\\
    \midrule
    Clone groups $|C|$ & 37 & 88 & 82 & 207\\
    Type-3 clone groups $|C^\mathit{T3}|$ & 21 & 21 & 65 & 107\\
    RQ~1: $|C^\mathit{T3}| / |C|$ & 0.56 & 0.23 & 0.79 & 0.52\\
    \midrule
    Faulty clone groups $|C_F|$& 16 & 5 & 37 & 58\\
    Faulty type-3 clone groups $|C^\mathit{T3}_F|$ & 7 & 1 & 2 & 10\\
    RQ~2: $|C^\mathit{T3}_F| / |C^\mathit{T3}|$ & 0.33 & 0.05 & 0.03 & 0.17\\
    \midrule
    Type-3 clones $|I|$ & 46 & 43 & 146 & 235\\
    Modified type-3 clones $|I_M|$ & 24 & 19 & 67 & 110\\
    Simultaneously modified type-3 clones $|I_\mathit{MS}|$ & 14 & 17 & 62 & 93\\
    RQ~3.1: $|I_\mathit{MS}| / |I_M|$ & 0.58 & 0.89 & 0.92 & 0.85\\
    \midrule
    Consistently fixed type-3 clone groups $|C_X|$ & 4 & 1 & 0 & 5\\
    RQ~3.2: $|C_X| / |C^\mathit{T3}_F|$ & 0.57 & 1.00 & 0 & 0.5\\
    \midrule
    Faulty type-2 clone groups $|C^\mathit{T2}_F|$ & 9 & 4 & 35 & 48\\
    Non-faulty type-3 clone groups $|C^\mathit{T3}_\mathit{NF}|$ & 14 & 20 & 63 & 97\\
    $|C^\mathit{T2}_F \rightarrow C^\mathit{T3}_\mathit{NF}|$ & 9 & 4 & 35 & 48\\
    RQ~3.3: $|C^\mathit{T2}_F \rightarrow C^\mathit{T3}_\mathit{NF}| / |C^\mathit{T2}_F|$ & 1 & 1 & 1 & 1\\
    \midrule
    Mean length of type-3 clones (in units) & 60 & 62 & 78\\
      Mean length of faulty type-3 clones (in units) & 50 & 39 & 83\\
    \bottomrule
  \end{tabular}
  \label{tab:results}
\end{table}

The development of the TFC started tightly coupled with the development of system A. Salient modules of the TFC were implemented as immediate contributions to system A. The co-existence of system A with a then immature TFC might have fostered cloning and as a consequence the occurrence of type-3 clones. This may be particularly true due to the missing baseline configuration of the TFC.

\highlightt{Answer to RQ~1:}{On average, every second clone group is a type-3 clone group. Therefore,
type-3 clones are a substantial part of all clones.}

\subsection{Type-3 Clones with Documented Faults (RQ~2)}

We found documented faults in 58 clone groups overall. Ten of these were type-3 clone groups. Interestingly,
while system C had the most faulty clone groups (37), system A had the most faulty type-3 clone groups (7).
Hence, in our small sample the relationship from faulty clone groups to faulty type-3 clone groups is not linear.
This could be an indication that some other factors, such as the developers' awareness of clones, play a role.

The ratio of faulty type-3 clone groups in relation to all type-3 clone groups is on average 17~\%. Because of the
discussed imbalance between faulty clone groups and faulty type-3 clone groups, this ratio varies strongly from
3~\% in system C to 33~\% in system A. Again, this could be an indication that the developers of systems B and C were
more aware of the clones and, hence, introduced less inconsistencies which represent faults.

Another explanation is that systems B and C, with low ratios of faulty type-3 clones, were particularly quality-checked because of their complex and computation-centric application logic. Also, systems B and C shared release cycles of about three months in contrast to system A with monthly releases.

Potentially, the ratio of faulty type-3 clone groups could be higher, because we only analysed documented faults.
There still might be faults in the code that have not been detected so far (as observed by Juergens et al.~\cite{Juergens:2009hy}).
Yet, we wanted to concentrate on faults that had an actual effect and led to failures.

\highlightt{Answer to RQ~2:}{On average, 17~\% of all type-3 clone groups contained a documented fault. The range
is from 3~\% to 33~\%. Therefore, type-3 clones do contain documented faults but not a high ratio of them. The differences between the cases suggest that other factors play a role.}

\subsection{Developers' Awareness of Type-3 Clones (RQ~3)}

In RQ~1, we established that type-3 clones are interesting to investigate. In RQ~2, we found that type-3 clones contain
documented faults as well as that there are differences in how many and what share of type-3 clone groups contain
documented faults. Previous research suggests that developer awareness of these clones plays a role. 
Therefore, we investigate
this awareness more closely. We analysed four different indications of this awareness which we discuss in the following.

\subsubsection{Simultaneous Maintenance of Type-3 Clones (RQ~3.1)}

For the analysis of simultaneous maintenance of type-3 clones, we did not look at the clone groups but at the individual 
clones (also called \emph{clone instances}) in
relation to their clone groups. Overall, we detected 235 type-3 clones in all three systems. Systems A and B had roughly
the same number of type-3 clones with 46 and 43, respectively. System C had the most type-3 clone groups and, accordingly,
the most type-3 clones: 146. Of these type-3 clones, 110 have been modified in the timeline we investigated. In all three systems
roughly half of the clones were modified. 

Developers changed most of these modified clones simultaneously with the other clones in the same clone group. Overall,
85~\% of the modifications in type-3 clones were simultaneous modifications. There is again a large imbalance in this ratio.
While for system B and C it is around 90~\% (89~\% and 92~\%), for system A it is 58~\%. Hence, the developers of system A
changed the type-3 clones much less often at the same time. This supports our previous theory that the awareness of developers
of the existence of clones plays a role in the ratio of faulty type-3 clone groups. Less simultaneous modification indicate a lesser
awareness. Yet, in general, the developers modified clones often simultaneously and, hence, seem to have been aware of the clones in 
their code.

\highlightt{Answer to RQ~3.1:}{Overall, 85~\% of the modifications to type-3 clones were done simultaneously. Hence, developers
seem to be highly aware of the clones. There are differences between systems, however, which suggest differences in the clone awareness of developers.}

\subsubsection{Consistent Fixes (RQ~3.2)}

The second indication for clone awareness we investigated was whether documented faults in type-3 clones were fixed
consistently by the developers. We found few faulty type-3 clone groups in general. Therefore, the numbers are not
particularly insightful. Yet, of the few faulty type-3 clone groups, 50~\% were fixed overall over time. The one faulty
group in system B was fixed. The developers did not remove the faults in the groups in system C. Of the 7 faulty groups
in system A, the developers fixed 4. 

A clear general conclusion is difficult. The 57~\% fix ratio for system A supports the hypothesis that its developers were
not highly aware of the clones. Yet, in system C none were fixed, while it had the highest ratio of simultaneously modified 
type-3 clones. We believe that the low number of faults distorts the results.

An explanation could be that system A with a particularly high ratio of faulty type-3 clones and a low ratio of synchronously fixed type-3 clones makes extensive use of Eclipse application perspectives. An Eclipse application perspective describes a certain configuration of views and editors. In system A, 7 different perspectives are defined for complementary user tasks. In particular, the perspectives contain very different editors and views from a \emph{user} perspective. They do share, however, a very similar code base for standard \emph{programming} tasks such as, for example, tabular presentations of TFC domain objects. The development of independent application perspectives requires little interaction by the different developers in charge. Code reuse by means of type-3 clones, however, seems to be particularly fault-prone due to this lack of developer communication and the resulting delayed fault fixes. This may indicate the usefulness of a recommender system that notifies developers of the remaining faulty clone instances.

\highlightt{Answer to RQ~3.2:}{The developers fixed consistently overall 57~\% of all faulty type-3 clone groups. The low number of
documented faults in clone groups prohibits clear conclusions.}

\subsubsection{Faulty Type-2 Clone Evolution (RQ~3.3)}

The third indication for developer awareness of clones we investigated was whether the inconsistencies in type-3 clones
resulted from fault fixes in type-2 clones. We detected 48 type-2 clone groups with documented faults. This ranges from
only 4 groups in system B to 35 groups in system C. 

By going from the fault detection forward in time, we found that all of the type-2 clone groups with a documented fault
evolved to type-3 clone groups without documented faults. Therefore, the ratio of such an evolution is 1 for all systems.
In other words, almost half of the type-3 clone groups came into existence by fault fixes in type-2 clone groups but for
none of these does another documented fault exist. This indicates that there is an awareness for clones by the developers
of all three systems. Otherwise, there should be occurrences of incomplete fixes, i.e.\ that a fix was not made to all
clone instances where it was needed.

\highlightt{Answer to RQ~3.3:}{All type-2 clone groups with documented faults evolved to type-3 clone groups with
not further documented faults. This indicates that the developers were aware of the clones and made the fixes to
all necessary instances.}

\subsubsection{Developers' View (RQ~3.4)}

The central codes we have derived from the interview transcripts are \emph{General clone awareness}/\emph{No
general clone awareness}, \emph{No specific clone awareness}, \emph{No clone check while bug fixing}, 
\emph{Clone warning while developing}, \emph{Common code ownership} and \emph{Discussion about co-changes}.
They allow us to describe the differences in clone awareness in the three cases. Table~\ref{tab:central-codes} shows
which of these codes describe which cases.

\begin{table}[htb]
  \caption{Assignment of central codes to cases}
  \centering
\begin{tabular}{l c c c}
\textbf{Code} & \textbf{A} & \textbf{B} & \textbf{C} \\
\hline
General clone awareness & & x\\
No general clone awareness & x & & x\\
No specific clone awareness & x & & x\\
No clone check while bug fixing & x & & x\\
Clone warning while developing &  & x \\
Common code ownership & & & x\\
Discussion about co-changes & & & x\\
\hline
\end{tabular}%
 \label{tab:central-codes}
\end{table}

The codes describe the three distinct profiles of the three cases. In case A, we did not find general awareness of
the concept of code clones and also in the project, there was no specific handling of code clones. During
bug fixes, there was no explicit analysis whether there are copies that would need to be changed as well.
Simultaneous changes are probably caused by other means of impact analysis in the project.

Case B is quite the opposite: There is a general awareness of the concept and potential problems of
code clones. Yet, during the project, code clones have not been discussed in detail. But there was a
check for clones during bug fixes with the \emph{duplicate code} warning by the static analysis tool 
\emph{Checkstyle}.\footnote{\url{http://checkstyle.sourceforge.net}}

In case C, we did not find general clone awareness either. Also there was no specific awareness and means
for analysis during bug fixes in the project. Yet, in the project, there is an emphasis on common code ownership and intensive discussions
on the code. In particular, there are discussions before a change about its consequences and what needs to
be changed as well. The developers seem to identify clones which need to be changed in this way.

Finally, we also defined the code \emph{Wish for clone warning while developing} capturing that in the
two cases that did not have clone warnings during development (A and C), the developers expressed
the wish to have this functionality. Both cases had duplication analysis with \emph{SonarQube}\footnote{\url{http://www.sonarqube.org}}
automatically performed with the build of the system. Those were described as too late. The information
is needed before or while performing a change.

\highlightt{Answer to RQ~3.4:}{In case A, there was neither general nor specific awareness of clones
during changes. In case B, there was general clone awareness and the tool Checkstyle showed warnings
of similar code in the IDE. In case C, there was neither general nor specific clone awareness. Yet, through
common code ownership and intensive communication before changes, other code location that need to 
be changed are identified. In all cases, the developers preferred warnings during development to clone results
together with an automated build.}

\subsubsection{Summary}

In summary, we found that the overwhelming majority of modifications to type-3 clones was done simultaneously
in the corresponding clone group. There were clear differences between the systems, however. The developers 
fixed the majority of faulty type-3 clone groups in all systems, and all fixed type-2 clone groups evolved to type-3
clone groups with no further documented faults.

Overall, we interpret these results such that developers took care to check the effects of changes and were able
to identify all locations where changes are necessary. Otherwise, we would see more faults from incomplete fixes.
Yet, there was a difference in the awareness of the clones. In the simultaneous modifications, we 
observe a difference between system A on the one side and systems B and C on the other. This could be an indication
for a difference in the level of awareness which, in turn, could explain the difference in the number of faulty clone groups
in the results for RQ~2. The developer interviews support this conjecture: While C could find similar code locations
by communication between the developers and B by general awareness and usage of a tool, A was not that aware
of clones. This is consistent with the quantitative results.

\highlightt{Answer to RQ~3:}{Developers of all three systems performed simultaneous updates to clones. Yet, the level
of awareness seems to be differing. A lower level of awareness could be an explanation for more faulty type-3 clone groups.}

\subsection{Length of Clones and Faults (RQ~4)}

Finally, we wanted to investigate the influence of the length of clones on the faultiness. We first analysed the
correlation between the length of clones (in units) and whether they contain a fault or not. Tab.~\ref{tab:correlation}
shows the results of an analysis with Spearman's correlation coefficient. The correlation (and effect size) is weak and 
not statistically significant.

\begin{table}[htb]
  \caption{The Results of Correlations Between Length of Clones and Faults}
  \label{tab:correlation}
\renewcommand{\arraystretch}{1.2} 
 \centering
 \begin{tabular}{ll r r}
 
\multicolumn{2}{l}{} & \multicolumn{1}{c }{Clone Length} & \multicolumn{1}{c}{Faults}  \\
\hline
\multirow{3}[6]{*}{Clone Length} & Spearman's rho Correlation & 1.000     & 0.268  \\
       & Sig. (2-tailed) & \multicolumn{1}{l}{} & 0.120  \\
       & N     & 35  & 35  \\
\hline
\end{tabular}%

\vspace{ 0.5 em}
\end{table}

Therefore, we have no empirical support for the assumption that longer clones are more likely to contain
faults. Hence, we do not further investigate the hypotheses but accept our null hypothesis: 
\emph{There is no difference in the length of clones between clone groups with a fault and clone groups without a fault.}

\highlightt{Answer to RQ~4:}{The length of clones do not influence their faultiness.}

\subsection{Evaluation of Validity}

\subsubsection{Construct validity} 
The manual analysis, we performed here, could have introduced problems. We only double-checked random samples.
Because we start from the latest versions of the source files, we inevitably miss clones created somewhere in the 
history and which are removed later on. We do not have data to quantify this threat.

\subsubsection{Internal validity}
Our manual false-positive removal could be wrong. We double-checked only random clone groups.
Regarding the configuration of ConQAT, the results of clone detection with conservative parameter settings (to increase precision) are shown 
in Tab.~\ref{tab:conservative} and with liberal settings (to increase recall) in Tab.~\ref{tab:liberal}. We found that the liberal approach delivers 
far higher numbers of clones.  Yet, in our manual analyses, they contained a lot of false positives. Therefore, we chose
the conservative approach for the further analysis and found no further false positives.

\begin{table*}[htb]
  \caption{Clone detection with the liberal approach}
  \centering
\begin{tabular}{cccccccc}

\textbf{Project} & \textbf{Minlength} & \textbf{Error} & \textbf{Gap Ratio} & \textbf{Runtime} & \textbf{kLOC} & \textbf{Clone LOC} & \textbf{Clone Count}  \\
 \hline
A     & 10    & 10    & 0.25  & 58s   & 253   & 25.443 & 981  \\
 
B     & 10    & 10    & 0.25  & 58s   & 332   & 49.2  & 1.545  \\ 
C     & 10    & 10    & 0.25  & 112s  & 454   & 47.8  & 2.244  \\
\hline
\end{tabular}%
 \label{tab:liberal}
\end{table*}

\begin{table*}[htb]
  \caption{Clone detection with the conservative approach}
  \centering
\begin{tabular}{ccccccccc}
 
\textbf{Project} & \textbf{Minlength} & \textbf{Error} & \textbf{Gap Ratio} & \textbf{Runtime} & \textbf{kLOC} & \textbf{Clone LOC} & \textbf{Clone Count}  \\
\hline
A     & 20    & 10    & 0.25  & 52s   & 253   & 7.6   & 143  \\
 
B     & 20    & 10    & 0.25  & 42s   & 332   & 17.7  & 352  \\
 
C     & 20    & 10    & 0.25  & 97s   & 454   & 15.6  & 382  \\
 \hline
\end{tabular}%
 \label{tab:conservative}
\end{table*}

A further threat to validity is that the three systems have been developed with a new issue-tracking system. Therefore, 
earlier issues could not be analysed. Accordingly, the amount of change sets does not fully represent the revision history.
Nevertheless, we believe the analysed history gives a good insight into the evolution of the clones and faults.

\subsubsection{External Validity} 
The study was only carried out on three relatively young industry systems that have thus also smaller version histories. Although 
all systems are written in Java and perform different functions, the results are fairly consistent over the various projects. As the
variety in software development processes, programming languages and application domains is huge, it is possible that
in other contexts, the numbers can vary considerably. Yet, we believe that the results should be comparable to other medium-sized
software systems written in languages similar to Java.

\section{Conclusions and Future Work}

\subsection{Summary of Conclusions}
In the three investigated industrial closed-source software systems, we found that on average, 
half of all clone groups are type-3 clones. Hence, they are a common phenomenon
interesting for further analysis.
Of all type-3 clone groups, only 17~\% contain documented faults. This is not negligible 
but type-3 clones are not a major source of faults in these systems. Nonetheless, the faultiness
was differing over the three systems.

We found a potential reason for this difference in faultiness is the awareness of the
developers of these clones. We found that most of the type-3 clones are modified simultaneously. 
Half of the faults in type-3 clones were fixed consistently. This suggests that overall the developers in our cases
have been successful in changing clones together when needed which led to the small
number of faults in type-3 clones.

The system with the highest rate of faulty type-3 clone groups, however, had also the lowest share 
of simultaneously changed clone groups. This was supported by the qualitative analysis that showed
that the developers of this system also had the least awareness of clones. This can be an
explanation for the higher number of faults.

Furthermore, all type-2 clones with documented faults in our systems evolved into fixed type-3 clones. 
This suggests that often type-3 clones are created to fix faults and do not introduce new faults or miss 
to fix faults.

Finally, there is only a weak correlation, which is not statistically significant, between clone length and 
faultiness. Therefore, clone length does not to be an important factor for future studies.

\subsection{Relation to Existing Evidence}

We found exactly the same rate of type-3 clones as Juergens et al.~\cite{Juergens:2009hy} (based on
the same detection technique). Hence,
it seems that consistently about half of all clone groups are type-3 clones. Also the rate of faulty type-3
clones is consistent. Juergens et al.\ found with 0.15 an only slightly smaller rate. Hence, it seems that
analysing documented faults or asking developers directly does not change the result. The result from
Bettenberg et al.~\cite{Bettenburg:2012fs} for faulty inconsistencies is with a maximum of 4~\% considerably
lower but they only looked at faults at the release level. Hence, it could still be in harmony with our results.

We cannot directly
compare the results for unintentional type-3 clones as Juergens et al.\ calculated that for the clone groups
and this study analysed the individual clones. Yet, Juergens et al.\ found that 28~\% of the clone groups
were unintentionally inconsistent while we found that 85~\% of the clones were changed simultaneously.
Hence, the tendency is in both cases that developers seem to be aware of the majority of type-3 clones.

G\"ode and Koschke~\cite{Gode:2011ct} concluded that ``only 14.8\% of all changes to clones are unintentionally
inconsistent.'' Thummalapenta et al.~\cite{Thummalapenta:2010bv} found that more than 70~\% of clone groups in their
analysed systems ``were either consistently changed or they underwent an independent evolution''. We can
support these findings as we found that 85~\% of type-3 clones were simultaneously modified.

Rahman, Bird and Devanbu~\cite{rahman10} investigated the relationship between faults and clones
in general. They found that cloned code does contribute \emph{less} to faults than non-cloned code and that
clones are not a major source of faults. We focused on type-3 clones and found that they often occur as
a result from fixing faults in type-2 clones. Hence, we can support the claim that clones are not a major
source of faults.

In the end, there does not seem to be such a large contradiction between the different studies but in the
interpretation. There are faults related to inconsistent changes in type-3 clones. Such faults are usually
not the majority of all faults in a system. Yet, the awareness of developers of these clones seems to play a role.

\subsection{Impact/Implications}

As long as developers are aware of clones and type-3 clones in particular, clones do not seem to be
a strong source of faults. We see three implications from this:
\begin{enumerate}
  \item Although it is a small number, type-3 clones do contain faults. It might still be interesting to use the inconsistencies
           between type-3 clones as a hint for finding faults (similar as in \cite{Juergens:2009hy} or \cite{Li:2006ip}). The efficiency of such an 
           approach in comparison to other fault detection approaches needs to be investigated.
  \item As several studies confirmed now that faultiness is not a strong argument for considering cloning to be a bad smell,
           researchers need to concentrate more on the effects of the size increase caused by cloning. There are some first
          analyses (e.g.\ fix effort in \cite{rahman10}) but no comprehensive investigations.
  \item Awareness of clones to handle them consistently seems to be a prerequisite for the low effect on faultiness of
           clones. Keeping track of these change dependencies in form of clones causes cognitive load on the developers.
           This suggests that clone detection and management tools that support developers in being aware of these
           dependencies can be helpful~\cite{Thummalapenta:2010bv}. Yet, the effect of cloning on cognitive load needs to be investigated.
\end{enumerate}

\subsection{Limitations}
This study covers an underrepresented area in studies so far: industrial closed-source software systems. Nevertheless, we only
investigated three systems from one company. Although our results fit well to most existing studies, replications
with other systems and companies in other domains are necessary to better establish the findings.

Furthermore, as we only looked at documented faults and not the individual inconsistencies, we could not find
faults not discovered otherwise so far. In that sense, we complement especially \cite{Juergens:2009hy}. This
could distort the findings, but our intuition was that faults not discovered after years of usage might not be so
interesting after all.

\subsection{Future Work}
As mentioned above, replications of this study in other contexts would help to increase confidence in the findings.
We plan to approach more industrial partners to conduct similar studies on their code bases. To be fully
explanatory, however, we will need a controlled experimental design.

Furthermore, in all three cases, the developers mentioned that the most useful way to present clones to them
would be directly in the IDE. We work on extensions to IDEs that support the awareness of developers of clones in code
fragments they are working on without putting unnecessary cognitive load and distractions on them. 


\section*{Acknowledgment}
We would like to thank the developers at TWT who allowed us to investigate their projects and in
particular the three project leads who took the time for the interviews.




\balance

\bibliographystyle{IEEEtran}

\end{document}